\documentclass[conference]{IEEEtran}

\usepackage{amsmath}
\usepackage{amsthm}
\usepackage{cite}
\usepackage{graphicx}
\usepackage{multirow}
\usepackage{url}
\usepackage{color}
\usepackage[normalem]{ulem}

	\title{Maximizing Aggregator Profit through Energy Trading by Coordinated Electric Vehicle Charging}
\author{\IEEEauthorblockN{James J.Q. Yu, Junhao Lin, Albert Y.S. Lam, and Victor O.K. Li}
\IEEEauthorblockA{Department of Electrical and Electronic Engineering\\
The University of Hong Kong\\
Pokfulam Road, Hong Kong\\
Email: \{jqyu, jhlin, ayslam, vli\}@eee.hku.hk}}

\begin{document}
	\maketitle
	
	\begin{abstract}
		Due to the increasing concern for greenhouse gas emissions and fossil fuel security, electric vehicles (EVs) have attracted much attention in recent years. EVs can aggregate together constituting the vehicle-to-grid system. Coordination of EVs is beneficial to the power system in many ways. In this paper, we formulate a novel large-scale EV charging problem with energy trading in order to maximize the aggregator profit. This problem is non-convex and can be solved with a centralized iterative approach. To overcome the computation complexity issue brought by the non-convexity, we develop a distributed optimization-based heuristic. To evaluate our proposed approach, a modified IEEE 118 bus testing system is employed with 10 aggregators serving 30 000 EVs. The simulation results indicate that our proposed distributed heuristic with energy trading can effectively increase the total profit of aggregators. In addition, the proposed distributed optimization-based heuristic strategy can achieve near-optimal performance.
	\end{abstract}

	\section{Introduction}
	
	\IEEEPARstart{W}{ith} the increasing concern for greenhouse gas emissions, electric vehicles (EVs) are expected to reach a significant market share in the near future. With the emerging vehicle-to-grid (V2G) technologies, EVs can potentially help alleviate the security concern for the supply of fossil fuels and mitigate the power network instability caused by the highly penetrated intermittent renewable energy generations \cite{Tabari2013}. However, the network security and economic operation can be significantly adversely influenced by the uncoordinated charging behaviors of a large number of EVs  \cite{PieltainFernandez2011}. By efficiently utilizing the system capacity, coordinated charging strategies can reduce the possible adverse impacts on the power system \cite{Han2010}. In addition, other merits can also be obtained, such as reducing the total operational cost and mitigating the variability of the renewable energy sources \cite{Yao2013}. Hence, coordination of EVs is beneficial to the power system in many different perspectives.
	
A large population of EVs can be clustered into aggregators to facilitate the coordination of charging behaviors. 
Coordinated EV charging within one aggregator has been studied extensively in recent years. Previous research generally focused on the grid structure with aggregators managing a large number of EVs. Based on where the charging decisions are made, the methodologies can be classified into two types: centralized and distributed methods. In the centralized methods, EVs need to send their parking and battery information to the associated aggregator, which then decides how each individual EV should be charged in a centralized manner. The decision is mostly driven by maximizing the aggregator profit \cite{Han2010,Sortomme2011}, minimizing EV owners' power cost \cite{Wu2012,Vandael2015}, and achieving power balance \cite{Hoog2015}. 
	On the other hand, aggregators in the distributed methods send power pricing scheme and related information to EVs. The EVs then utilize their own knowledge to establish their charging plans, which are later delivered to the aggregator. Usually this process repeats until an agreement or equilibrium is reached. As the charging strategies are developed by the EV, most existing efforts employ the energy costs incurred for the EV owners as the performance metric \cite{He2012}. User utilization ratio \cite{Wen2012} and power balance \cite{Gan2013} are also studied as metrics. In general, all these mainly focus on the coordinated charging behavior within a single aggregator.
	
There is few efforts studying the interactions among multiple EV aggregators, e.g., \cite{Qi2014,Shao2016}. One common feature is the utilization of a hierarchical architecture where aggregators coordinate with others under the control of the system operator. While this architecture efficiently reduces the computational burden of developing coordinated charging plans, direct interactions among aggregators are generally not considered.

    In this paper, we focuses on maximizing the aggregators' total profit through coordinated EV charging when considering multiple aggregators connected to the grid. Different from the previous work, we investigate the possibility of achieving Inter-aggregator Energy Trading (IET) to further exploit profits from trading energy directly among aggregators. There exists previous research investigating the possible energy trades among EVs, e.g., \cite{Wu2012a} and \cite{Kim2013}, and they demonstrated improved total costs of EVs when compared with those models without energy trading. Therefore it is likely that energy trading among aggregators is more profitable with direct aggregator interactions. In addition, we consider the Locational Marginal Pricing (LMP) in this work to manage the system congestion cost to further increase the profit.

The remainder of this paper is organized as follows. Section II presents the system framework of our proposed approach. We formulate the aggregator profit maximization problem and propose a centralized coordinated charging strategy in Section III. Section IV demonstrates an optimization-based heuristic approach for the problem. In Section V we provide a case study to illustrate the performance of our developed approach. Finally we conclude this paper in Section VI.

	\section{System Architecture}

In this work, we adopt a typical hierarchical EV charging architecture \cite{Qi2014}, which is composed of three components: system operator (SO), aggregators, and plug-in EVs. The main purpose of introducing aggregators is to simplify the system model and unbundle the electric loads, e.g., EVs, from the power network infrastructure \cite{Hansen2015}. By collecting the charging information from the clustered EVs and pricing information from SO, each aggregator can develop its own optimal charging schedule such that the aggregators' profits are maximized while the various EV charging requirements are all satisfied. SO manipulates the real-time power price to balance power consumption of each bus in the power grid to alleviate power congestion.

We consider EVs as dispatchable power appliances whose charging rates can be adjusted by the corresponding aggregators by taking the system requirements and economic consideration into account \cite{Wen2012}. Similar to \cite{Lam2015}, upon the arrival of an EV, we assume that its battery capacity and current state of charge (SoC) can be obtained by its aggregator via appropriate vehicular communication techniques. In addition, the departure time of the EV is also assumed to be available to its aggregator. 
    We allow EVs to perform early departure if needed. In such case, a penalty can be imposed to the EV owner \cite{Olivares2014}. 
	
	In our proposed framework, EVs are categorized into two classes based on their respective capabilities: uni-directional, i.e., charging from the grid to vehicles, or bi-directional, i.e., both charging from and discharging to the grid. Enabling power flow from EVs to the grid can effectively increase the profit by exploiting the power price fluctuations \cite{Han2010,Sortomme2011,Suekrue2015}. To encourage V2G bi-direction operation, aggregators may compensate the EV owners' battery aging loss by offering lower charging cost or free battery replacement plan \cite{Sortomme2012}.
	
	
    The objective of IET is to effectively utilize the discharging power from aggregators for the benefit of aggregator profit.
    The aggregators typically purchase power from SO for charging at a price, and sell power to SO at a lower price. Meanwhile, instead of trading with SO, the discharging aggregators can trade with other charging aggregators at a better price, between the SO power buying and selling prices. For instance, suppose that Aggregator A1 is charging power $P_1 >0$ from the grid with power purchasing price $C^\text{ch}_1$ and Aggregator A2 is discharging power $P_2 < 0$ to the grid with power selling price $C^\text{dch}_2$.
Then the total cost for the aggregators is $P_1C^\text{ch}_1+P_2C^\text{dch}_2$.
Suppose that A2 desires to sell $P^\text{trade} > 0$ to A1 at $C^\text{trade}$. 
The power cost for A1 becomes $(P_1 - P^\text{trade})C^\text{ch}_1 + P^\text{trade}C^\text{trade}$ and A2's is $(P_2 + P^\text{trade})C^\text{dch}_2 - P^\text{trade}C^\text{trade}$, rendering a total cost at $P_1C^\text{ch}_1+P_2C^\text{dch}_2 + (C^\text{dch}_2 - C^\text{ch}_1)P^\text{trade}$. 
Therefore the trades between A1 and A2 can decrease the cost, thus increase the aggregators' total profit when $C^\text{dch}_2>C^\text{ch}_1$.

	\section{Aggregator Profit Maximization Problem}\label{sec:centralized}
	
	
In this section, we formulate the aggregator profit maximization problem with IET.
We employ model predictive control to develop charging control strategies of the immediate time slot, taking future EV departures and power price profile into account. Specifically, the problem is defined over a finite time-horizon $\mathcal{T}=\{t_q|t_q=t_0+q\Delta t, q=0,1,\cdots,q^\textit{max}\}$. 
	Information in the current time slot $t_0$ is considered accurate while the future information may be imprecise. Since the solution of the problem, i.e., the overall charging strategy, should be jointly optimal over the entire $\mathcal{T}$, the implemented charging strategy of the current time slot will also contribute to the total profit maximization over $\mathcal{T}$, instead of one single time slot. The problem shall be solved whenever EVs require charging operations, rendering the process online. All symbols used in this paper and their meanings are listed in Table \ref{tbl:nomenclature}.
    
	\begin{table*}
		\caption{Sets, Parameters, and Variables Used}
		\centering
		\label{tbl:nomenclature}
		\begin{tabular}{l|l||l|l}
			\hline
			Parameter & Description & Parameter & Description. \\
			\hline
			$\mathcal{A}$ & Set of aggregators in the system. & $P^\text{line}_l$ & Maximum power flow on line $l$. \\
			$A_i$ & $i$-th aggregator in the system. & $G^\text{min}_{s}$,$G^\text{max}_{s}$ & Minimum and maximum generation output for the \\
			$\mathcal{V}_i$ & Set of Electric Vehicles (EVs) served by $A_i$. & & generator on bus $s$. \\
			$V_{i,j}$ & $j$-th EV served by $A_i$. & $t$ & Current time slot. \\
			$V_{i,u}$ & $u$-th unidirectional charging EV served by $A_i$. & $R_t$ & The total aggregator profit at $t$. \\
			$\mathcal{T}$ & Time horizon of optimization. & $P_i$ & Total charging power of $A_i$. \\
			$\mathcal{T}_{i,j}$ & Parking time of $V_{i,j}$. & $P_{i,j}$ & Charging power of $V_{i,j}$. \\
			$t_0$,$t_q$ & The current and the $q$-th future time slot. & $\alpha_{i,j}$ & Binary indicator for whether $V_{i,j}$ is still parked after \\
			$q^\text{max}$ & The total number of future time slots. & & the registered departure time. \\
			$\Delta t$ & Duration of each time slot. & $S_{i,j}$ & SoC of $V_{i,j}$. \\
			$t^\text{dpt}_{i,j}$ & The registered departure time of $V_{i,j}$. & $S^\text{rmin}_{i,j}$ & Minimum reserved State-of-Charge (SoC) of $V_{i,j}$. \\
			$C_i^\text{ch}$ & Power charging price from the power grid at $A_i$. & $G_s$ & Generation output for the generator on bus $s$. \\
			$C_i^\text{dch}$ & Power discharging price from the power grid at $A_i$. & $P_s$ & Aggregator injection on bus $s$. \\
			$C^\text{V}$,$C^\text{U}$ & The base contract price for bi-directional and uni- & $C_{i,j}$ & Charging fee of $V_{i,j}$ when the registered parking time \\
			& directional vehicle-to-grid (V2G) charging enabled EVs. & & is $\mathcal{T}_{i,j}$. \\
			$\Delta C^\text{V}$,$\Delta C^\text{U}$ & The marginal price decrease for bi-directional and & $C^\text{trade}$ & Power trading price among aggregators. \\
			& uni-directional V2G charging enabled EVs. & $C_i$ & Power trading price from the power grid at $A_i$. \\
			$T^\text{V}$,$T^\textit{U}$ & Parking time for minimal charging fee of bi-directional & $P^\text{trade}_i$ & Trade power of $A_i$. \\
			& and uni-directional V2G charging enabled EVs. & $P^\text{spl}_i$ & Total available aggregator power supply in the trade \\
			$P^\text{ch}_{i,j}$,$P^\text{dch}_{i,j}$ & Maximum charging and discharging rate of $V_{i,j}$. & & at $C_i$. \\
			$\eta^\text{ch}_{i,j}$,$\eta^\text{dch}_{i,j}$ & Charging and discharging efficiency of $V_{i,j}$. & $P^\text{dmd}_i$ & Total available aggregator power demand in the trade \\
			$E_{i,j}$ & Battery capacity of $V_{i,j}$. & & at $C_i$.\\
			$S^\text{min}_{i,j}$,$S^\text{max}_{i,j}$ & Minimum and maximum State-of-Charge (SoC) of $V_{i,j}$. & $P^\text{cap}_i$ & Trade capacity at $C_i$. \\
			$S^\text{req}_{i,j}$ & Required SoC on departure of $V_{i,j}$. & $C_s(G_s)$ & Power generation cost function for the generator on bus \\
			$N$ & Number of buses in the power system. & & $s$ when generating $G_s$ power. \\
			$L$ & Number of lines in the power system. & $\lambda$,$\mu_l$ & Lagrangian multipliers.\\
			$D_{s}$ & Inelastic load on bus $s$. & $\gamma^\text{min}_s$,$\gamma^\text{max}_s$ & Lagrangian multipliers.\\
			$F_{l-s}$ & Generation shift factor from bus $s$ to line $l$. & & \\
			
			\hline
		\end{tabular}
	\end{table*}
	
	
	In the problem, a set of $m$ aggregators $\mathcal{A}=\{A_1, A_2, \cdots, A_i, \cdots, A_m\}$ are considered. Aggregator $A_i$ serves $n_i$ EVs at a time denoted by $\mathcal{V}_i=\{V_{i,1}, V_{i,2}, \cdots, V_{i,j}, \cdots, V_{i,n_i}\}$. The objective function for time slot $t$ is mathematically formulated as follows:\footnote{For the sake of simplicity, the symbol $t$ for time is omitted in equations (1)--(3), (6), and (8)--(11), when no confusion may be caused.}
    
    \vspace{-0.5cm}
	{\small
    \begin{align}\label{eqn:objective}
		R_t & = \underbrace{\sum_{A_i\in\mathcal{A}}\sum_{V_{i,j}\in\mathcal{V}_i}\alpha_{i,j}P_{i,j} C_{i,j}}_{\text{Charging Income}} + \underbrace{\sum_{A_i\in\mathcal{A}}\sum_{V_{i,j}\in\mathcal{V}_i}(1 - \alpha_{i,j})P^\text{ch}_{i,j}C_{i,j}}_{\text{Penalty Income}} \nonumber \\
		& - \underbrace{\sum_{A_i\in\mathcal{A}}(\sum_{V_{i,j}\in\mathcal{V}_i}\alpha_{i,j}P_{i,j} - P^\text{trade}_i)C_i}_{\text{Marginal Energy Cost}} 
	\end{align}}where
\begin{equation}
	C_i = 	
	\begin{cases}
	C_i^\text{ch}  & \text{if } \sum_{V_{i,j}\in\mathcal{V}_i}\alpha_{i,j}P_{i,j} - P^\text{trade}_i \geq 0,\\
	C_i^\text{dch} & \text{otherwise},
	\end{cases}
\end{equation}
which means that the charging price is employed if the aggregator draws power from the grid. Otherwise the discharging price is used. The income of aggregators is composed of two parts: charging and penalty incomes. The first term in Eq. \eqref{eqn:objective} is the charging fee imposed on the EVs, which is arbitrarily formulated as follows:\footnote{More complicated contract formulations are available in practice and can be easily incorporated into the proposed optimization problem.}

\vspace{-0.4cm}
{\small
\begin{equation}\label{eqn:chargingfee}
		C_{i,j} = 	
		\begin{cases}
		C^\text{V} - \Delta C^\text{V} \min\{|\mathcal{T}_{i,j}|/T^\text{V},1\} & \text{bidirectional charging},\\
		C^\text{U} - \Delta C^\text{U} \min\{|\mathcal{T}_{i,j}|/T^\text{U},1\} & \text{unidirectional charging},
		\end{cases}
\end{equation}}

\vspace{-0.4cm}
\noindent where the bidirectional charging allows both charging and discharging behaviors, and the unidirectional charging supports charging only. The second term in Eq. \eqref{eqn:objective} is the penalty income (parking fee) imposed on the EVs for any late departure. This income can be neglected in residential area scenarios. The third term in Eq. \eqref{eqn:objective} is the cost of the consumed power purchased from SO. The cost is firstly generated using day-ahead energy prices, then taking LMP into consideration in optimization of the later time slots. If the charging power is negative, the corresponding aggregator will perform V2G energy selling operation and this term will become negative.
	
	
	Besides real time profit generation, aggregators can also utilize the dispatchable characteristics of EVs to delay the charging process to further increase the profit. This results in a mult-time-slot joint optimization as follows:
    
    \vspace{-0.5cm}
{\small\begin{IEEEeqnarray}{l}\IEEEyesnumber\label{eqn:optimization}\IEEEyessubnumber*
		\hspace{0.0cm}\underset{P_{i,j,t},P^\text{trade}_{i,t}}{\text{maximize}} \sum_{t\in\mathcal{T}}R_t\hspace{0.3cm}\text{subject to} \label{fit:1}\\ 
		P^\text{dch}_{i,j} \leq P_{i,j,t} \leq P^\text{ch}_{i,j}, \forall t\in \mathcal{T},\label{con:1}\\
		0 \leq P_{i,u,t} \leq P^\text{ch}_{i,u}, \forall t\in \mathcal{T},\label{con:2}\\
		P_{i,j,t}\Delta t + (t^\text{dpt}_{i,j} - t - 1)P^\text{ch}_{i,j} \geq (S^\text{req}_{i,j} - S_{i,j,t})E_{i,j}/\eta^\text{ch}_{i,j},\label{con:3}\\
        \IEEEeqnarraymulticol{1}{r}{\forall t\in \mathcal{T},\nonumber}\\
		P_{i,j,t}\Delta t\eta^\text{ch}_{i,j} \leq (S^\text{req}_{i,j} - S_{i,j,t})E_{i,j}, \forall t\in \mathcal{T},\label{con:5}\\
		\sum_{A_i\in\mathcal{A}}P^\text{trade}_{i,t} = 0, \forall t\in \mathcal{T},\label{con:6}\\
		P^\text{trade}_{i,t}\sum_{V_{i,j}\in\mathcal{V}_i}P_{i,j,t} \geq 0,\forall t\in \mathcal{T}, \label{con:7}\\
		\sum_{V_{i,j}\in\mathcal{V}_i}P_{i,j,t} - |P^\text{trade}_{i,t}| \geq 0, \forall t\in \mathcal{T}. \label{con:8}
	\end{IEEEeqnarray}}
    
    \vspace{-0.35cm}
\hspace{-0.4cm}	Eq. (\ref{con:1}) imposes rigid upper bounds for charging power of all EVs online. 
    Eq. (\ref{con:2}) further limits the uni-directional charging EVs to perform charging operations only  during the whole parking process.
	Eq. (\ref{con:3}) ensures that the current charging power of an EV is feasible only if the battery can be charged to the required SoC when the charging powers of all succeeding time slots are set to the maximum rate. This constraint is considered when the optimization horizon $\mathcal{T}$ does not contain the departure time $t^\textit{dpt}_{m,i}$, i.e., $\max\{\mathcal{T}\}<t^\textit{dpt}_{m,i},\exists m,i$.
    Otherwise, an alternative constraint is considered instead of Eq. (\ref{con:3}), as:
	\begin{equation}\label{con:4}
		\sum_{t\in\mathcal{T}}P_{i,j,t}\Delta t\eta^\text{ch}_{i,j} \geq (S^\text{req}_{i,j} - S_{i,j,t})E_{i,j}.
	\end{equation}
    This constraint guarantees that EVs will be charged to their required SoC on departure. Other practical strategies to prevent insufficient charging, e.g., penalties on aggregators, can be easily implemented in our model.
    Eq. (\ref{con:5}) prevents the battery from being over charged by limiting the current charging rate.
    Eqs. (\ref{con:3}) and (\ref{con:5}) cooperate to manipulate the charging rates of EVs to satisfy the SoC constraints.
    Eq. (\ref{con:6}) ensures that the amount of energy bought and sold in the aggregator trading market are equal. Non-convex constraint Eq. (\ref{con:7}) prevents the aggregators, that are consuming power from the power grid, from trading energy to other aggregators, and vice versa. As $P^\text{trade}_{i,t}$ and $\sum_{V_{i,j}\in\mathcal{V}_i}P_{i,j,t}$ are independent and can be both positive and negative. Eq. (\ref{con:8}) imposes that the aggregators cannot sell more energy than they generate through discharging behaviors, and cannot buy more energy than they request to fulfill the EV charging demand. 
    
    When given constant power prices $C_i$'s, the optimization problem formulated in Eq. \eqref{eqn:optimization} can develop optimal EV charging profiles for maximizing aggregator profit. However, the optimized charging powers of each aggregator $P_i=\sum_{V_{i,j}\in\mathcal{V}_i}\alpha_{i,j}P_{i,j}$ may change the power flow of the grid, resulting in changing $C_i$'s. Therefore, we also employ an Optimal Power Flow (OPF) based LMP optimization together with Eq. \eqref{eqn:optimization} to develop $C_i$'s subject to aggregators' changing $P_i$ values.
	
	
	Upon the receipt of charging power requests from the aggregators, SO performs OPF to route the power flow. In our proposed framework, an LMP strategy is employed for congestion control \cite{Wang2014a}. As we focus on the real time power dispatch of the aggregators for EV charging, it is assumed that the Unit Commitment problem has been solved and all generators considered are online \cite{Yao2013}. The SO level OPF problem is formulated as a classical OPF problem \cite{Li2014},
 	where the LMP of buses are optimized with the partial derivative of the Lagrangian of OPF (see \cite{Li2014,Wang2014a} for elaborations).
	
	
	\begin{figure}
		\centering
		\includegraphics[width=0.85\linewidth]{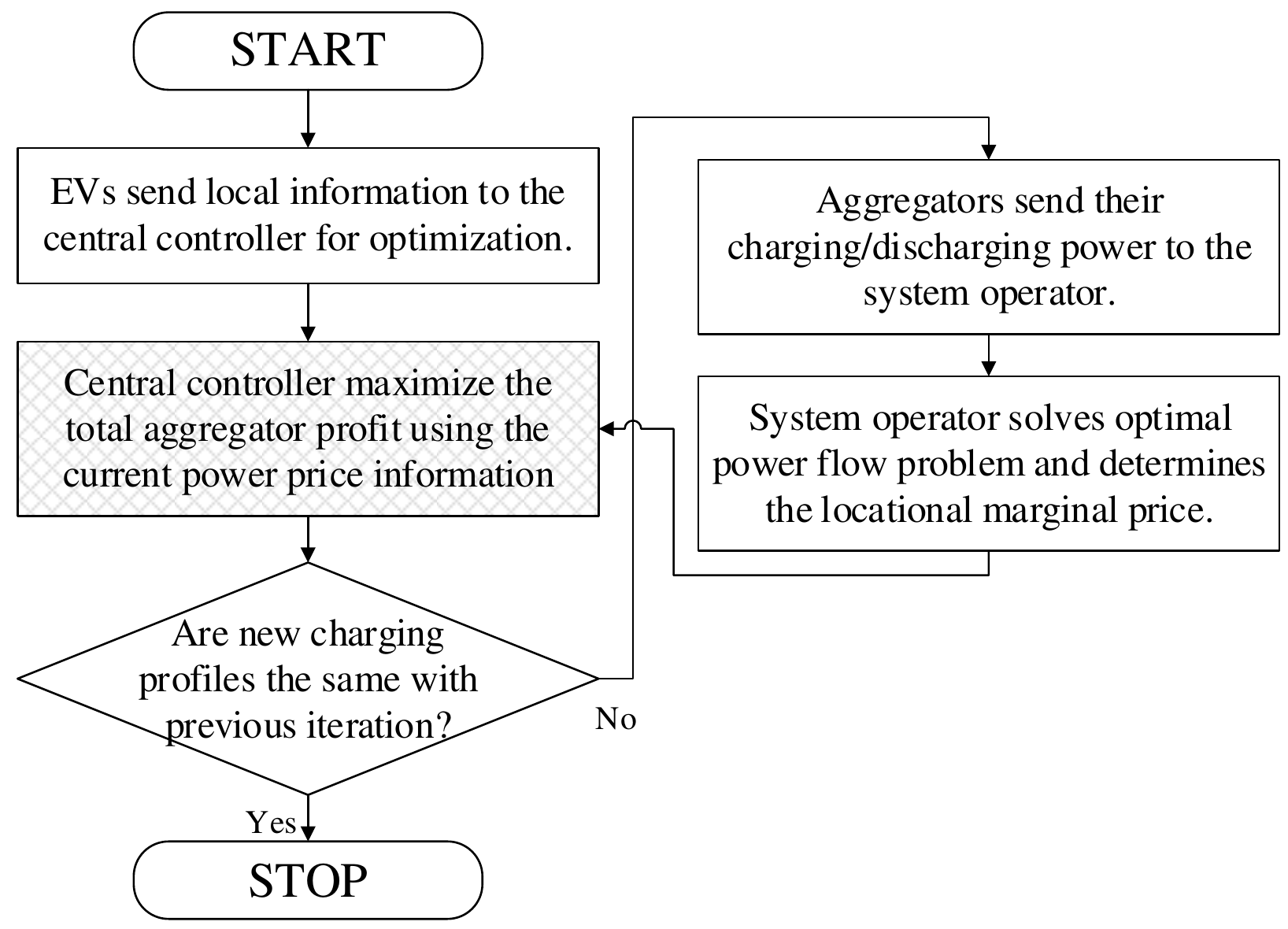}
        \vspace*{-3mm}
		\caption{Flow chart for the proposed centralized coordinated charging strategy for aggregator profit maximization.}
        \vspace*{-4mm}
		\label{fig:iteration_flowchart}
	\end{figure}
	
	\begin{figure}
		\centering
		\includegraphics[width=0.85\linewidth]{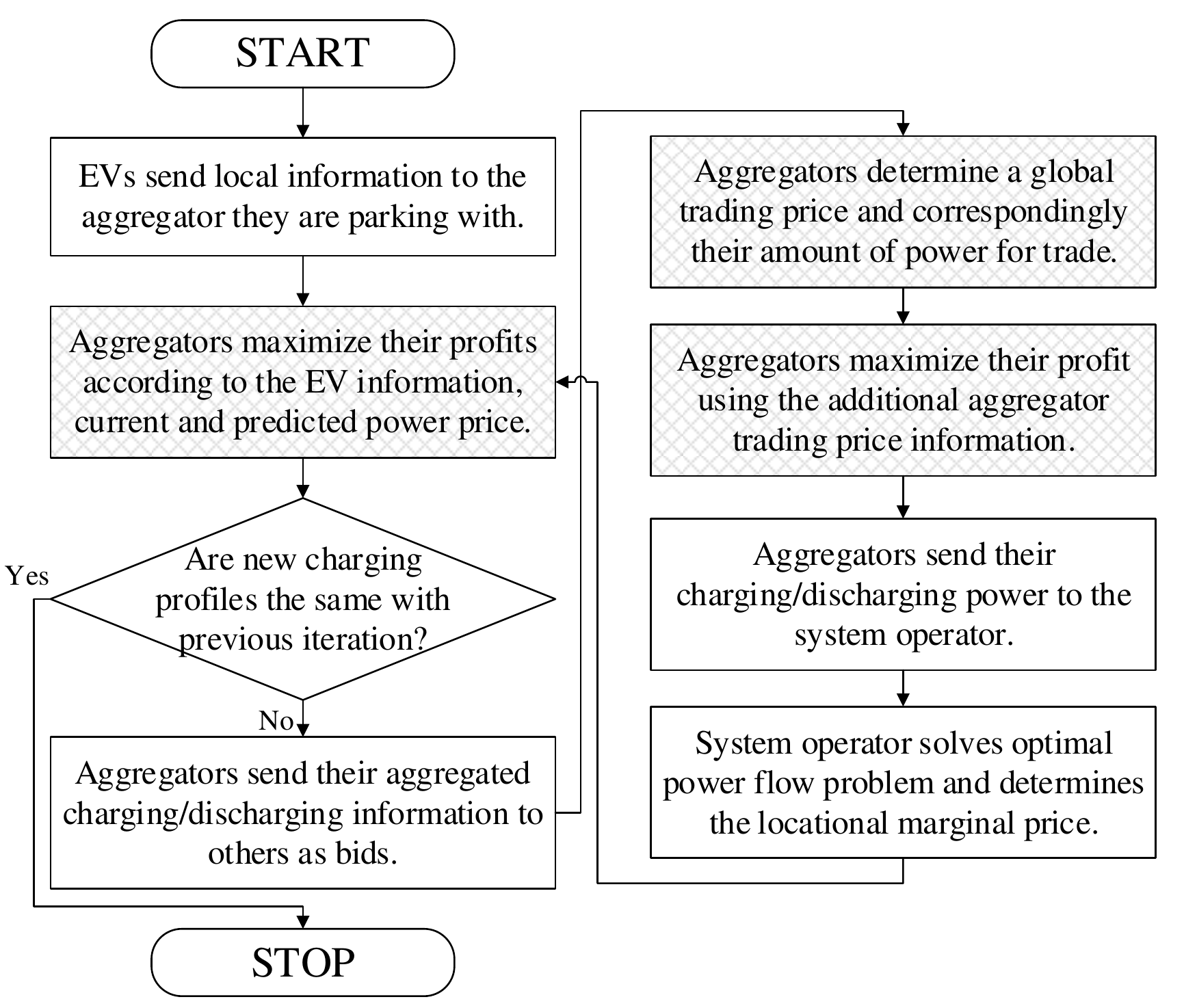}
        \vspace*{-3mm}
		\caption{Flow chart for the proposed optimization-based heuristic coordinated charging strategy for aggregator profit maximization.}
        \vspace*{-6mm}
		\label{fig:flowchart}
	\end{figure}
    
    Optimization problem \eqref{eqn:optimization} and OPF-based LMP optimization together compose the aggregator profit maximization problem. This problem aims to develop optimal EV charging plans to maximize the aggregator profit. It can be solved in an iterative manner as depicted in Fig. \ref{fig:iteration_flowchart}. Problem \eqref{eqn:optimization} and OPF are optimally solved alternately until an equilibrium is reached, where a stable IET-enabled EV charging profile is developed. At the beginning of each time slot, Eq. \eqref{eqn:optimization} is solved to generate optimal charging strategies for all EVs in the system, using the day-ahead energy price.
After the scheduled aggregated charging powers are optimized, they are reported to the SO for calculating the new LMP. Utilizing these new prices, the central controller solves Eq. \eqref{eqn:optimization} again, and this process repeats until an optimal charging strategy is developed.

	\vspace*{-0.1cm}
	\section{Distributed Aggregator Heuristic}
	\vspace*{-0.1cm}
		
	The aggregator profit maximization problem develops optimal EV charging profiles for maximizing aggregator profit. However, the problem can become intractable when there is a large number of EVs to schedule. As Eq. \eqref{eqn:optimization} is a non-convex problem, existing solvers may encounter efficiency issues when solving large problem instances. In order to address this problem, we propose a distributed heuristic algorithm to solve Eq. \eqref{eqn:optimization}. The heuristic-enabled aggregator profit maximization strategy is shown in Fig. \ref{fig:flowchart}. 
We basically divide the shaded step in Fig. \ref{fig:iteration_flowchart} (which constitutes the non-convexity) into the three shaded sub-steps in Fig. \ref{fig:flowchart}. They respectively correspond to the three steps of the proposed heuristic: aggregator profit optimization (shaded sub-step on the left in Fig. \ref{fig:flowchart}), power trading (top-right), and supply-demand balancing (middle-right).
	
	\vspace{-0.3cm}
	\subsection{Aggregator Profit Optimization}
	\vspace{-0.2cm}

	At the beginning of each optimization iteration, all online EVs send their vehicle and battery information, including the maximum charging rate, scheduled departure time, and battery size and current SoC, to their corresponding aggregator. With this information, each aggregator maximizes its profit independently using a modified formulation of Eq. (\ref{eqn:objective}):

\vspace{-0.3cm}
	{\small\begin{align}\label{eqn:aggregator_profit}
	R_{i, t} & = \sum_{V_{i,j}\in\mathcal{V}_i}\alpha_{i,j}P_{i,j}C_{i,j} + \sum_{V_{i,j}\in\mathcal{V}_i}(1 - \alpha_{i,j})P^\text{ch}_{i,j}C_{i,j} \nonumber \\
	& - (\sum_{V_{i,j}\in\mathcal{V}_i}\alpha_{i,j}P_{i,j} - P^\text{trade}_i)C_i - P^\text{trade}_iC^\text{trade}.
	\end{align}}
\vspace{-0.2cm}
	
	The major difference between Eqs. (\ref{eqn:objective}) and (\ref{eqn:aggregator_profit}) lies in the introduction of the trading cost term $P^\text{trade}_iC_\text{trade}$. $P^\text{trade}_i$ is set to zero at the beginning of the iteration, and set to either a constant value or to $\sum_{V_{i,j}\in\mathcal{V}_i}\alpha_{i,j}P_{i,j}$ based on the supply and demand equilibrium, which will be elaborated in Section IV-B. Therefore, Eq. \eqref{eqn:optimization} is transformed into
	\begin{IEEEeqnarray}{l}\label{eqn:aggregator_optimization}
		\underset{P_{i,j,t}}{\max} \sum_{t\in\mathcal{T}}R_{i,t}\hspace{0.3cm}\text{subject to } 
		\text{(\ref{con:1}), (\ref{con:2}), (\ref{con:3}) or (\ref{con:4}), (\ref{con:5}), \label{fit:3}}
	\end{IEEEeqnarray}
	to maximize each aggregator's profit. Eq. \eqref{eqn:aggregator_optimization} considers all EV constraints, and energy trading constraints Eqs. \eqref{con:6}--\eqref{con:8} are handled in the power trading heuristic algorithm.
	
	
	
	\subsection{Power Trading Heuristic}
	
	When all aggregators have finished their profit maximization processes, they broadcast their $P_i$ values to the others in the aggregator trading market in the form of bids. In this step, the aggregators utilize all the power bids to perform a modified second-price auction \cite{Myerson1981}.
	When submitting bids, each aggregator calculates its $P_i$ from the result of Eq. (\ref{eqn:aggregator_optimization}), and places a bid in the form of $(P_i, C_i)$ pair. A positive $P_i$ makes $C_i=C^\text{ch}_i$, and a negative $P_i$ makes $C_i=C^\text{dch}_i$. The bids are then broadcast to the others and the auction starts in the aggregator trading market when all bids have been placed and announced.
	
	After all bids are generated, the auction evaluates the available charging (demand) and discharging (supply) power for trade at all possible trading prices $C^\text{trade}$, whose values are selected from all $C_i$ values in the bids:
	\begin{align}\label{eqn:calculate_supplydemand}
	P^\text{spl}_i = \sum_{A_k\in\mathcal{A}^\text{spl}_i} P_k, 
	P^\text{dmd}_i = \sum_{A_k\in\mathcal{A}^\text{dmd}_i} P_k,
	\end{align}
	where
	\begin{align}
	\mathcal{A}^\text{spl}_i = \{A_k\in\mathcal{A}| P_k<0, C_k\leq C^\text{trade}\}, \label{eqn:selling}\\
	\mathcal{A}^\text{dmd}_i = \{A_k\in\mathcal{A}| P_k>0, C_k\geq C^\text{trade}\}. \label{eqn:buying}
	\end{align}
	The trading capacity for $C^\text{trade}$ is accordingly calculated by
	\begin{equation}
	P^\text{cap}_i = \min\{-P^\text{spl}_i,P^\text{dmd}_i\}\times C^\text{trade}.
	\end{equation}
Consequently, the value of $C^\text{trade}$ is set to the optimal $C_i$ value which yields the largest $P^\text{cap}_i$.
    
After determining the trading price, another round of optimization Eq. (\ref{eqn:aggregator_optimization}) is conducted to utilize the additional profit contributed by the trading behavior. The generated charging profile is considered final and reported to SO for LMP calculation.

	\vspace{-0.1cm}
	\section{Case Studies}
	\vspace{-0.1cm}
	
	
	
	We employ the IEEE 118 Bus system \cite{PowerSystemsTestCaseArchive-ref} to assess the profit maximization performance of our proposed approach. 10 aggregators are installed in the system on Buses 7, 14, 17, 28, 44, 58, 72, 84, 97, and 115. 
	The settings of the system parameters are presented in Table \ref{tbl:parameters}. The power price information is acquired from the PJM \cite{PJMEnergyMarketInformation-ref} data, and the aggregators are assigned with the prices of different buses.
	
	In the test system, 30 000 EVs are accommodated. We consider two models of vehicles, namely the Tesla Model S AWD-85D (Telsa Model S) and Nissan Leaf 2014 model (Nissan Leaf). We assume 60\% of all EVs in the system are Tesla Model S EVs with 85 kWh batteries and 34 kWh/100 miles energy expenditure performance, and the remaining 40\% are Nissan Leaf EVs with 24 kWh batteries and 31.64 kWh/100 miles energy expenditure performance. The maximum charging rates for these two models are 22 kW and 6.6 kW, respectively, and the maximum discharging rates are also set to the same values \cite{Nguyen2014}. Among all EVs in the system, 80\% EVs enabled bi-directional V2G operations. 5\% of the EVs will perform a late departure up to a maximum of one hour. For practical situations, these parameters can be manipulated by the EV owners. The European Commission Strategic Energy Technologies Information System reported a mobility survey on the driving and parking patterns of European car drivers \cite{Pasaoglu2012}. The results of the survey are utilized to formulate the EV driving dynamics. Similar methods have also been adopted in the literature, e.g, \cite{Yao2013}.
	
	
	
	\begin{figure*}
		\centering
		\includegraphics[width=0.8\linewidth]{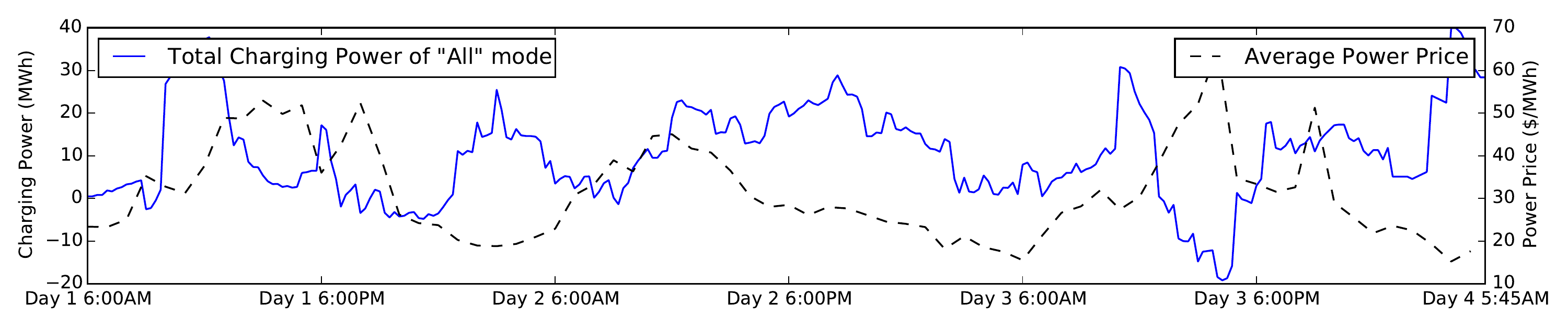}
        \vspace*{-2mm}
		\caption{Total charging loads for all time slots.}
        \vspace*{-4mm}
		\label{fig:charging}
	\end{figure*}
	
	
	We perform simulations on a horizon of 72 hours, and the length of each time slot is 15 minutes. Thus the total profit of all 288 time slots are combined as the performance metric.
    As Problem \eqref{eqn:optimization} cannot be solved directly in a timely manner, we first investigate the performance of the distributed heurisric-enabled aggregator profit maximization strategy proposed in Section IV, whose optimality will be studied later. The distributed solution is labeled ``All'' for ease of demonstration. In addition, four other variants are considered to demonstrate the relationship of Eq. \eqref{eqn:aggregator_optimization} and LMP calculation, and their contributions to the profit maximization performance. These four variants are developed by removing one or multiple components from ``All'':
	\begin{enumerate}
		\item The \textit{NoTrade} mode removes all possible trades between aggregators. This mode works like a V2G enabled version of \cite{Li2014}.
		\item The \textit{NoLMP} mode removes the LMP adaptation step. Energy trades among aggregators are allowed.
		\item The \textit{Planning} mode removes both trading steps and the LMP adaptation step. 
        This mode works similar to a V2G-enabled version of \cite{Suekrue2015}.
		\item The \textit{Greedy} mode performs the greedy charging strategy. Upon the arrival of an EV, it is charged at the maximum rate until the SoC requirement is met.
	\end{enumerate}
	
	\begin{table}
		\caption{Parameter Settings}
        \vspace*{-0.2cm}
		\centering
		\label{tbl:parameters}
		\begin{tabular}{cc|cc|cc}
			\hline
			Param & Value & Param & Value & Param & Value \\ \hline
			$q^\text{max}$ & 23 & $\Delta R^\text{V}$ & \$0.015/6hrs & $\eta^\text{ch}_{i,j}$,$\eta^\text{dch}_{i,j}$ & 0.9 \\
			$\Delta t$ & 15 minutes & $\Delta R^\text{D}$ & \$0.015/6hrs & $S^\text{min}_{i,j}$ & 0.0 \\
			$R^\text{V}$ & \$0.08/kWh & $T^\text{V}$ & 6 hours  & $S^\text{max}_{i,j}$ & 1.0 \\
			$R^\text{U}$ & \$0.10/kWh & $T^\text{U}$ & 6 hours & $S^\text{req}_{i,j}$ & 0.9 \\\hline
		\end{tabular}
    \vspace*{-0.5cm}
	\end{table}
	
	\begin{table}
		\caption{Profit Maximization Performance}
        \vspace*{-0.2cm}
		\centering
		\label{tbl:performance}
		\begin{tabular}{l|ccccc}
			\hline
			Mode & All & NoTrade & NoLMP & Planning & Greedy \\ \hline
			Profit(\$) & 134295.1 & 110269.8 & 118534.4 & 109750.6 & 75678.9 \\ \hline
		\end{tabular}
    \vspace*{-0.5cm}
	\end{table}
	
	Table \ref{tbl:performance} presents the total profits generated by the compared charging approaches. It can be easily seen that our proposed approach can significantly increase the aggregator profits, and both the aggregator trading and LMP adaptation have a positive influence on the profit maximization process when compared with the Greedy mode. In addition, the computational time of the proposed approach is also crucial as the algorithm is supposed to be implemented online. With 30 000 EVs in the system and 24 optimized time slots, each iteration of our proposed algorithm can be finished in 9.13 seconds. All time slots can be optimized within a maximum of six iterations, i.e., one time slot can be finished in one minute.
	
	
	In addition to the profit maximization performance comparison, we also investigated the optimality of our proposed approach in Section IV (``All'' mode) comparing with the original one in Section III, and the behavior of aggregators in response to the power price. The detailed results are presented in \cite{Yu2015a}, and we observe that ``All'' is only 3.7\% worse than the optimal solution of the convex-relaxed problem, and the gap between the true optimal and our approximation must be smaller. Fig. \ref{fig:charging} demonstrates the total charging power dynamics of our proposed approach with the changes on the average power purchasing price. We can observe that the charging power is mostly high when the price is relatively low. This shows the efficacy of the multiple time slot optimization in saving aggregators' power cost.  The impact of introducing the proposed strategy on the power system stability is also illustrated in \cite{Yu2015a}, and the conclusion can be drawn that adjacent buses to aggregators will be influenced more by the power line congestion, which is represented in the form of LMP changes.

    \vspace*{-0.1cm}
	\section{Conclusion}
    \vspace*{-0.2cm}

	In this paper we propose a coordinated charging problem considering IET for maximizing the profit of multiple aggregators. A model predictive control based centralized iterative approach is devised to find the optimal EV charging strategies for profit maximization. Considering the non-convexity nature of the optimization problem, we develop an aggregator-level coordinated charging heuristic to construct the EV charging schedules. To exploit the potential of employing energy trade among aggregators for profit maximization, we propose an auction-based heuristic to handle the trading details. In addition, the iterative approach utilized by our developed strategy can further adapt the EV charging schedules to the grid congestion cost. To validate the performance of the proposed approach, we employ a ten-aggregator system with 30 000 EVs for simulation. The system is installed on an IEEE 118 bus system, and the simulation is performed on a 72 hours time span. The charging schedule developed by our proposed approach can create more profit for the aggregators than the compared strategies. Moreover, the proposed distributed optimization-based heuristic is compared with the relaxed convex optimization problem. The result shows that our heuristic can achieve near optimal performance.
    \vspace{-0.2cm}
	\section*{Acknowledgment} \vspace{-0.2cm}
This research is supported in part by the Theme-based Research Scheme of the Research Grants Council of Hong Kong, under Grant No. T23-701/14-N.

\bibliographystyle{IEEEtran}
\bibliography{IEEEabrv,reference}
	
\end{document}